\documentclass[12pt,a4paper]{article}
\usepackage[affil-it]{authblk}

\usepackage{siunitx}

\usepackage{hyperref}
\usepackage{amsfonts}
\usepackage{amsmath}
\usepackage{subfig}
\usepackage{mathtools,slashed,tensor}
\usepackage{graphicx}
\usepackage{lineno}
\usepackage[font=scriptsize,labelfont=bf]{caption}
\usepackage{float}
\usepackage[top=2cm, bottom=2cm, left=3cm, right=3cm]{geometry}
\usepackage{siunitx}
\usepackage{changepage}
\usepackage{color}
\definecolor{darkgreen}{rgb}{0,0.35,0}

\def\be{\begin{equation}}
\def\ee{\end{equation}}

\def\bea{\begin{eqnarray}}
\def\eea{\end{eqnarray}}

\def\I{{_{\rm{I}}}}\def\II{{_{\rm{II}}}}

\usepackage{tikz}
\usetikzlibrary{matrix}
\usepackage{enumerate}

\def\Tr{\mathrm{Tr}}

\def\si{\sigma}

\def\1{{_{1}}}\def\2{{_{2}}}

\def\noHe0{:\;\!\!\;\!\!:H_e(0):\;\!\!\;\!\!:}
\def\noHm0{:\;\!\!\;\!\!:H_\mu(0):\;\!\!\;\!\!:}

\def\si{\sigma}

\def\1{{_{1}}}\def\2{{_{2}}}

\def\I{{_{\rm{I}}}}\def\II{{_{\rm{II}}}}

\usepackage{tikz}
\usetikzlibrary{matrix}
\usepackage{enumerate}

\providecommand{\keywords}[1]
{
  \small	
  \textbf{\textit{Keywords---}} #1
}

\def\HH{\mathsf{H}} 
\def\PP{\mathsf{P}} 

\def\BH{\mathrm{BH}} 
\def\Tr{\mathrm{Tr}}

\def\HH{\mathsf{H}} 
\def\PP{\mathsf{P}} 

\def\BH{\mathrm{BH}} 
\def\Tr{\mathrm{Tr}} 
\def\ii{\mathrm{i}}  

\newcommand{\ket}[1]{| #1 \rangle} 
\newcommand{\bra}[1]{\langle #1 |} 

\newcommand{\ucharles}{Faculty of Mathematics and Physics, Charles University, V Hole\v{s}ovi\v{c}k\'ach 2, 18000 Prague 8, Czech Republic}
\newcommand{\uach}{Instituto de Ciencias F\'isicas y Matem\'aticas, Universidad Austral de Chile, Casilla 567, 5090000 Valdivia, Chile}
\newcommand{\archimea}{Arquimea Research Center, Camino de las Mantecas, 38320, Santa Cruz de Tenerife, Spain}
\newcommand{\uwarsaw}{Faculty of Physics, Institute of Theoretical Physics, University of Warsaw, Ul. Pasteura 5, 02-093, Warsaw, Poland}

\date{}

\begin{document}

\title{Hunting Quantum Gravity with Analogs: the case of graphene}
\author[1]{Giovanni Acquaviva\thanks{\href{mailto:gioacqua@gmail.com}{gioacqua@gmail.com}}}
\author[2]{Alfredo Iorio\thanks{\href{mailto:alfredo.iorio@mff.cuni.cz}{alfredo.iorio@mff.cuni.cz}}}
\author[3,2]{Pablo Pais\thanks{\href{mailto:pais@ipnp.troja.mff.cuni.cz}{pais@ipnp.troja.mff.cuni.cz}}}
\author[4]{Luca Smaldone\thanks{\href{mailto:luca.smaldone@fuw.edu.pl}{luca.smaldone@fuw.edu.pl}}}

\affil[1]{\archimea}
\affil[2]{\ucharles}
\affil[3]{\uach}
\affil[4]{\uwarsaw}

\maketitle

\begin{abstract}
Analogs of fundamental physical phenomena can be used in two ways. One way consists in reproducing specific aspects of classical or quantum gravity, of quantum fields in curved space or of other high-energy scenarios, on lower-energy corresponding systems. The ``reverse way'' consists in building fundamental physical theories, for instance, quantum gravity models, inspired by the lower-energy corresponding systems. Here we present the case of graphene and other Dirac materials.
\end{abstract}


\section{Introduction}

Richard Feynman wrote beautiful and visionary pages on analogs, in a famous lecture titled ``Electrostatic Analogs'', available in~\cite{Feynman} (see also~\cite{Xons} for comments and discussions). There he explains \textit{how} it happens that different physical systems, among which a solid analogy can be established, are all described in a unified manner. In his now famous words, this happens because ``the same equations have the same solutions''. Therefore, if we have no access to certain regimes of system A, but they correspond to certain reachable regimes of the analogous system B,
we can perform experiments on system B, and establish results valid for system A.

In the final, and less known, part of the lecture, he ventures into a visionary attempt to explain \textit{why} this is so. This goes on till the thrilling hypothesis of the existence of \textit{more elementary constituents} than the ones we deem to be fundamental. All those systems, including electrostatics itself, are just different coarse-grained versions of one dynamics, even more fundamental than quantum electrodynamics. Amazingly, Feynman realizes that the physical properties of space itself play a crucial role in the identification of such fundamental objects (that he calls ``little $X$ons'' \cite{Feynman}). It is precisely when space itself, besides matter, is included as part of the emergent phenomenon, that these are also the conclusions of certain completely independent arguments of the contemporary quantum gravity (QG) \cite{bekensteinBound1,bekensteinBound2,bekensteinBound3,aischol,carroll}.

As for the field of \textit{gravity} analogs, see~\cite{Barcelo2005}, the seminal work is the 1981 paper of Bill Unruh~\cite{UnruhAnalog}, where he proposed to search for the experimental signatures of the Unruh effect~\cite{bill} and of the Hawking effect~\cite{haw0}, in a fluid-dynamical analog.

Due to our deeper theoretical understanding of these phenomena and to the higher experimental control of condensed matter systems, it is now becoming increasingly popular to reproduce that and other aspects of fundamental physics in analog systems. Examples include the Hawking phenomenon in Bose--Einstein condensates~\cite{MunozdeNova2019}, the Weyl symmetry~\cite{i2,weylgraphene} and the related Hawking/Unruh phenomenon on graphene~\cite{iorio2012,iorio2014,iorio2015}, gravitational and axial anomalies in Weyl semimetals~\cite{Gooth2017}, ``moir\'{e} gravity'' in bilayer graphene~\cite{MoireGravity1,MoireGravity2}), and more.

Let us mention, for instance, the beautiful example of supernova explosions simulated in the laboratory by plasma implosions induced by intense lasers. Both systems are examples of fluid dynamics, and the Euler equations are invariant under an inversion transformation, that is an arbitrary uniform expansion or contraction of the system. This symmetry is studied in cosmology, and allows to map an explosion problem to a dual implosion problem. In principle this duality allows the complete three-dimensional evolution of highly structured explosion ejecta to be modeled using a static target in an implosion facility. In~\cite{LOR2001} the maximal invariance group was determined to be the semi-direct product of the Galilei group with $SL(2,R)$ the latter containing time-translations, dilations, and the inversion. Those results had an important impact on the field. More details are in the contribution~\cite{UniverseLaser} to this Issue.

Gravity analogs are not limited to condensed-matter systems, as shown by heavy-ion collisions at high energy~\cite{Castorina2007,caiosa1,caio1}. In fact, hadron production in high-energy scattering processes is just a Unruh effect in Quantum Chromodynamics (QCD) \cite{Castorina2007}, with the Unruh temperature, $T_U = (\hbar a)/(2\pi c k_B)$, given by the hadronization temperature, $T_h \simeq \SI{160}{\mega\electronvolt}$, hence corresponding to an enormous acceleration, $a\simeq \SI{4.6e32}{\metre\per\second^{2}}$, that makes the effect very easy to detect. Here we explicitly write Planck constant, the speed of light and Boltzmann constant. More details on this approach are in the contribution~\cite{casatziorio} to this Issue.

Here we focus on the proposal of graphene as an analog of high-energy fundamental physics~\cite{i2,weylgraphene,iorio2012,iorio2014,iorio2015,ipfirst,ip3,reach}, based on the fact that its low-energy excitations~\cite{PacoReview2009} are massless Dirac pseudo-relativistic fermions (the matter fields $\psi$), propagating in a carbon two-dimensional honeycomb lattice. The emergent (long-wave limit) description of the latter is a surface (a portion of spacetime described by the ``emergent'' metric  $g_{\mu \nu}$). Such a behavior is shared by a wide range of materials, ranging from silicene and germanene through d-wave superconductors to topological insulators~\cite{wehling}. Each of those materials has its own peculiarities, which allow for further extensions of results obtained with graphene, and hence permit to explore a wider range of the high-energy target systems. Let us now give some details.

Despite those impressive advances in the highly active area of analog physics, there are still two milestones to reach. One is to fully understand the epistemic role of analogs in fundamental high energy physics, as not all theorists would agree that analogs are much more than mere \textit{divertissements}. In fact, experimental results obtained with analogs are not used as feedbacks for the target theories they are analogs of (see, e.g.,~\cite{Dardashti2016,Xons}). Another milestone would be a reliable definition of an analog black-hole (BH) entropy, or at least, of a quantum field theory (QFT)-like entanglement entropy that, in the presence of horizons, might serve the scope of setting-up some form of the second principle of BH thermodynamics.

Any progress in this direction would be truly important for the QG research. Having some results there, we could eventually be able to address the so-called {\it information paradox}, i.e., the apparent loss of information during BH evaporation, a question that, most probably, cannot be entirely solved via theoretical reasonings. See, e.g., \cite{Penrose1996,Hawking2004,Almheiri2013,Hooft2016,mannblack} for different points of view.

In fact, there is plenty of unreachable regimes in fundamental physics, starting from BHs, that we do know to exist, but that are not (easily or at all) reproducible in a laboratory. It is then of tremendous interest to establish solid criteria for such systems to correspond to other systems, within our reach, and to perform experiments on the latter to know of the former. On the other hand, when such correspondences are solidly established, why not inferring from the analog system the most intimate nature of the target system? For instance, if QG behaves like graphene (under certain conditions for graphene and for certain specific regimes of QG), and since we still do not know how QG really is, why not trying to guess the whole QG picture from what we learned of the partial overlap between the two sytems?

This is a less beaten track, but not a completely empty one. For instance, inspired by the findings of~ \cite{i2,weylgraphene,iorio2012,iorio2014,iorio2015,ipfirst,ip3,reach}, in~\cite{aischol} the authors propose the existence of fundamental, high-energy constituents underlying both matter and space, and that these, at our low energies, exist in an entangled state. This entanglement is there because both, matter and space, emerge from the dynamics of the same more fundamental objects, whose existence can be inferred from the celebrated upper bound on the entropy of any system, conjectured by Bekenstein~\cite{bekensteinBound1}. Quoting Feynman and paraphrasing Bekenstein, those objects are called ``$X$ons'' \cite{Xons}. If such a view is correct, even matter that we deem be fundamental, i.e., elementary, is in fact ``quasi-matter'', just like the massless quasi-particles, $\psi$, of graphene~\cite{PacoReview2009} owe their properties to the interaction with the lattice\footnote{Interestingly, there is a proposal called ``atoms of spacetime matter'' that could be closely related with this concept of Xons~\cite{Singh2021,Singh2022}.}. The most noticeable result of this ``quasi-particle picture'' \cite{aischol} is that the evaporation of a BH inevitably leads to an information loss, in the sense that, in general, there is a nonzero entanglement entropy associated to the final products of the evaporation. On the other hand, within the same picture, in~\cite{aismal} the authors describe BH evaporation from the point of view of the $X$ons. They see there that the Bekenstein bound~\cite{bekensteinBound1,bekensteinBound2,bekensteinBound3} can be an effect of the Pauli exclusion principle, and that a full unitary picture, leading to a complete recovering of the initial information, is only possible if one could track the evolution of those fundamental constituents.

The paper is organized in two large Sections and some concluding remarks. Section \ref{SecGraphene} is dedicated to graphene and Dirac materials (DMs) as analogs of high-energy fundamental physics. Section \ref{SecBekensteinXons} is dedicated to the QG that the latter research has inspired. Each large Section has many Subsections. As for Section \ref{SecGraphene}: Subsections \ref{SubSecFirstScale}, \ref{SubSecSecondScale} and \ref{SubSecWeylSymm} explain the main reasons why graphene is good at reproducing scenarios of fundamental physics; then Subsection \ref{SubSecRamifications} tells about old, new and future developments of this line of research, dedicating to each topic a brief Subsubsection; finally, Subsection \ref{SubSecExperiment} comments on the experimental search. As for Section \ref{SecBekensteinXons}: Subsection \ref{SubQuasiPartPict} introduces the quasi-particle picture in the QG context; then Subsections \ref{SubBHEvaporationI} and \ref{SubBHevaporationII} deal with BH evaporation as seen from the quasi-particles and as seen from the $X$ons, respectively; the last Susection \ref{SubSecEmergingSpace} comments on recent work on how (classical) space emerges from the underlying (quantum) dynamics of $X$ons during BH evaporation. Section \ref{SecConclusions} is dedicated to our concluding remarks and are a chance to point to future developments of the whole analog enterprise, in general, and of that based on graphene, in particular.

\section{Analog gravity on graphene}\label{SecGraphene}

Graphene is an allotrope of carbon. It is one-atom-thick; hence it is the closest to a two-dimensional object in nature. It was theoretically speculated about it~\cite{wallace,semenoff} and, decades later, it was experimentally found \cite{geimnovoselovFIRST}. Its honeycomb lattice is made of two intertwined triangular sub-lattices $L_A$ and $L_B$, see Fig.~\ref{honeycombpaper}. As is now well known, this structure is behind a natural description of the electronic properties of $\pi$ electrons\footnote{As the $\pi$ electrons do not participate in the stronger covalent $\sigma$ bonds, these electrons are not so attached to the carbon nuclei and are freer to ``hop'' from an atom to a neighbour one.} in terms of massless, $(2+1)$-dimensional, Dirac (hence, relativistic-like) quasi-particles.

\subsection{First scale, $E < E_\ell$: from the tight-binding to the Dirac Hamiltonian}\label{SubSecFirstScale}

\begin{figure}[H]
\centering
 \includegraphics[scale=0.40]{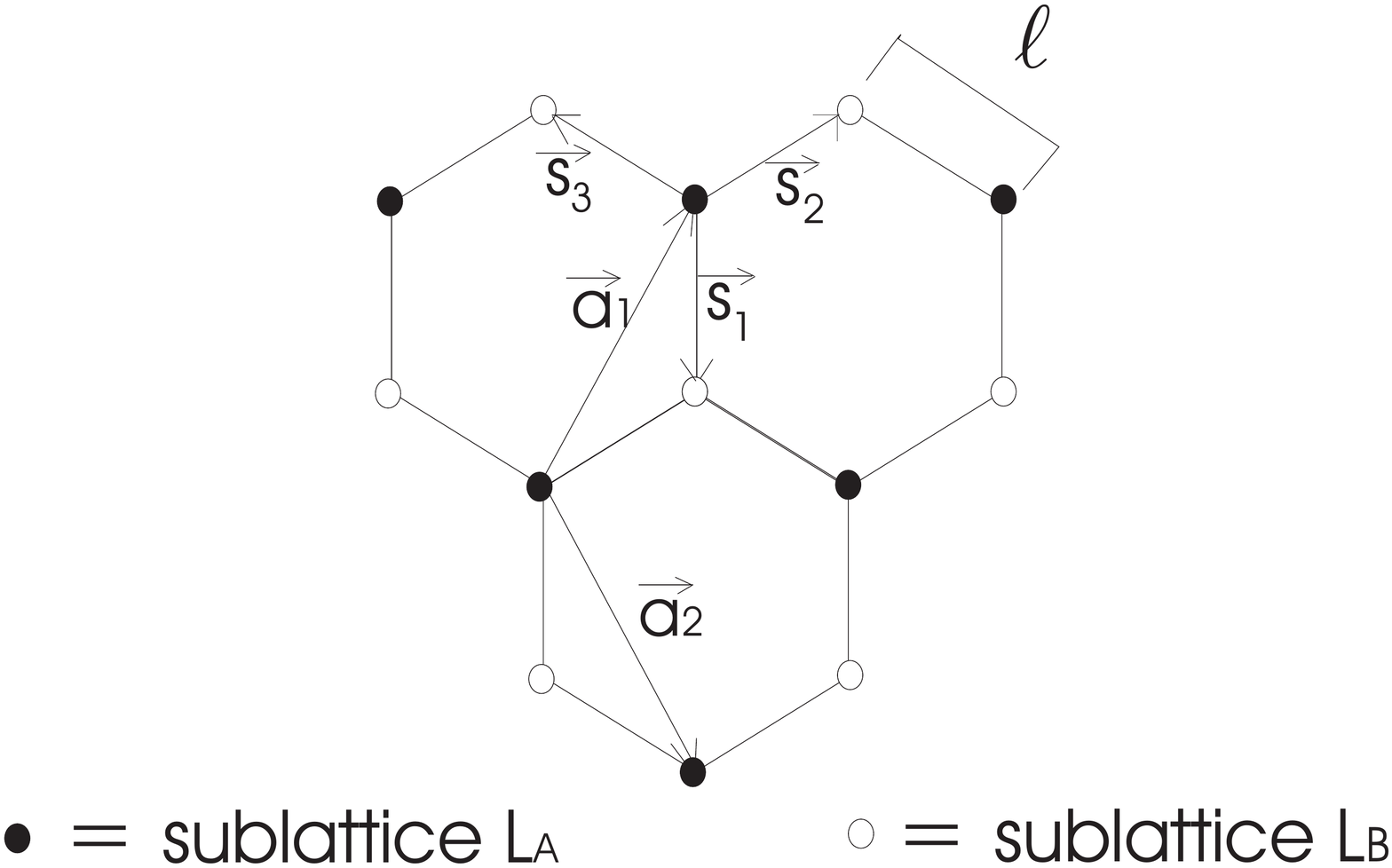}
\caption{The honeycomb lattice of graphene, and its two triangular sublattices $L_A$ and $L_B$. The choice of the basis vectors, $(\vec{a}_1, \vec{a}_2)$ and   $(\vec{s}_1, \vec{s}_2, \vec{s}_3)$, is, of course, not unique. Here we indicate the one used in \cite{iorio2015}. Figure taken from \cite{iorio2014}.}
\label{honeycombpaper}
\end{figure}

Such electrons, in the tight-binding low-energy approximation, are customarily described by the Hamiltonian (as here we use natural units, the reduced Planck constant is $\hbar = 1$)
\begin{linenomath}
\begin{equation}
    H = - \eta \sum_{\vec{r} \in L_A} \sum_{i =1}^3 \left( a^\dagger (\vec{r}) b(\vec{r} +\vec{s}_i)
    + b^\dagger (\vec{r} +\vec{s}_i) a (\vec{r}) \right) \label{primaH} \;,
\end{equation}
\end{linenomath}
where the nearest-neighbour hopping energy is  $\eta \simeq \SI{2.8}{\electronvolt}$, and $a,\, a^{\dagger}$ ($b,\, b^{\dagger}$) are the anti-commuting annihilation and creation operators, respectively, for the planar $\pi$ electrons in the sub-lattice $L_A$ ($L_B$), see Fig.~\ref{honeycombpaper}. All the vectors are bi-dimensional, $\vec{r} = (x,y)$, and, for the choice of basis vectors made in Fig.~\ref{honeycombpaper}, if we Fourier transform, $a(\vec{r}) = \sum_{\vec{k}} a(\vec{k}) e^{\ii\, \vec{k} \cdot \vec{r}}$, etc,  then $H = \sum_{\vec{k}} ( f(\vec{k}) a^\dagger (\vec{k}) b(\vec{k}) + {\rm h.c.})$, with
\begin{equation}
  f(\vec{k}) = - \eta \; e^{-\ii\, \ell k_y} \left( 1 + 2 \; e^{\ii\, 3 \ell k_y / 2} \cos(\sqrt{3} \ell k_x / 2) \right) \;,
\end{equation}
where $\ell\simeq\SI{1.4}{\angstrom}$ is the graphene honeycomb lattice length (see Fig.~\ref{honeycombpaper}). Solving $E(\vec{k}) = \pm |f (\vec{k})| \equiv 0$ tells us if, in the first Brillouin zone (FBZ), conductivity and valence bands touch and where. Indeed, this does happen for graphene, pointing to a gapless spectrum, for which we expect massless excitations to emerge. Furthermore, the solution is not a Fermi \textit{line} (the $(2+1)$-dimensional version of the Fermi surface of the $(3+1)$ dimensions), but instead, they are two Fermi \textit{points}, $\vec{k}^D_\pm = \left( \pm \frac{4 \pi}{3 \sqrt{3} \ell}, 0 \right)$. Even if the mathematical solution to $|f (\vec{k})| = 0$ has six points, only the two indicated are inequivalent \cite{PacoReview2009}.

\begin{figure}[H]
\begin{adjustwidth}{-5cm}{-5cm}
\centering
  \includegraphics[scale=0.80]{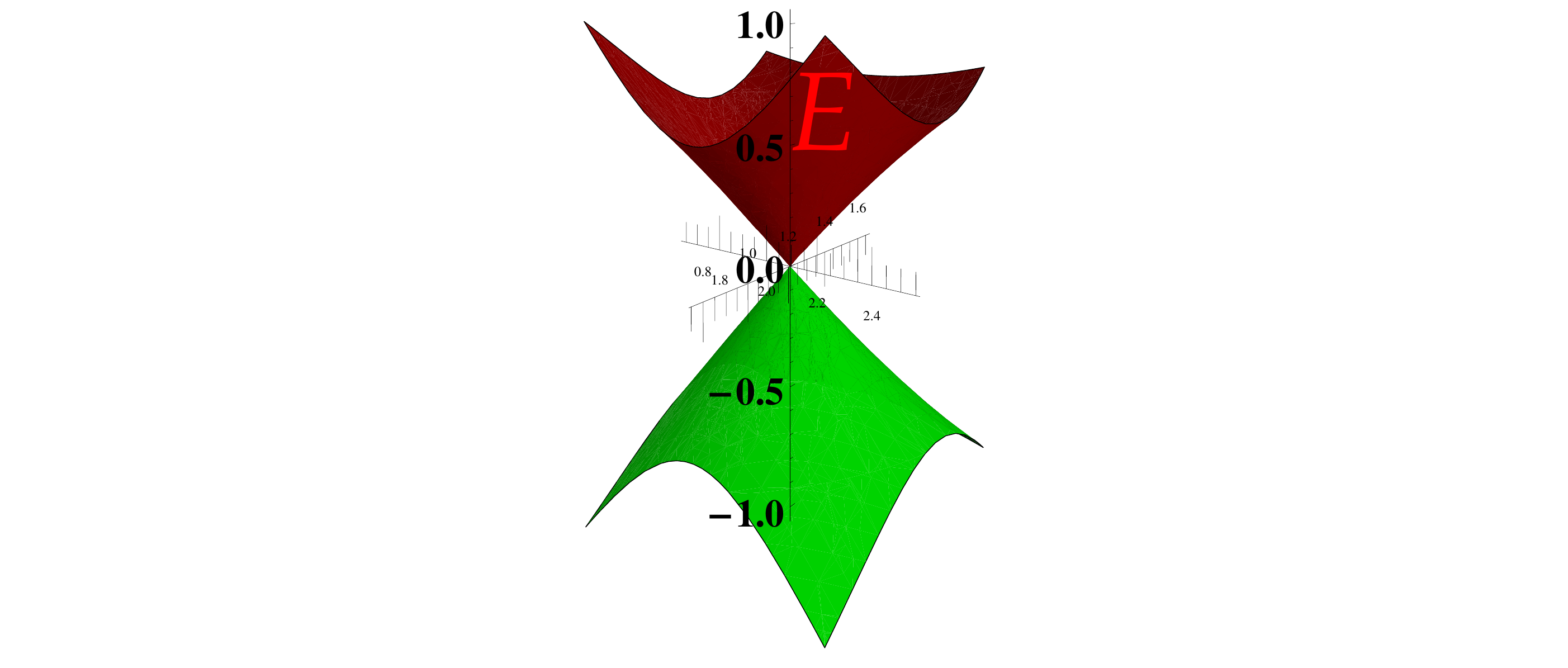}
\end{adjustwidth}
\caption{The linear dispersion relations near one of the Dirac points, showing the typical behavior of a relativistic-like system (the ``$v_F$-light-cone'' in $k$-space). Figure taken from \cite{iorio2015}.}
\label{lindisprel}
\end{figure}

The label ``$D$'' on the Fermi points stands for ``Dirac''. That refers to the all-important fact that, near those points, the spectrum is {\it linear}, as can be seen from Fig.~\ref{lindisprel}, $E_{\pm} \simeq \pm v_F |\vec{k}|$, where $v_F = 3 \eta \ell / 2 \sim c/300$ is the {\it Fermi velocity}. This behaviour is expected in a relativistic theory, whereas, in a non-relativistic system, the dispersion relations are usually quadratic.

If one linearizes around $\vec{k}^D_\pm$, $\vec{k}_\pm \simeq \vec{k}^D_\pm + \vec{p}$, then $f_+ (\vec{p}) \equiv f (\vec{k}_+) = v_F (p_x + i p_y)$,
$f_- (\vec{p}) \equiv f (\vec{k}_-) = - v_F (p_x - i p_y)$,  and $a_\pm (\vec{p}) \equiv a (\vec{k}_\pm)$,  $b_\pm (\vec{p}) \equiv b (\vec{k}_\pm)$. Therefore, the Hamiltonian (\ref{primaH}) becomes
\begin{eqnarray}
    H|_{\vec{k}_\pm} & \simeq & v_F \sum_{\vec{p}} \left(\psi_+^\dagger \vec{\sigma} \cdot \vec{p} \; \psi_+
    - \psi_-^\dagger \vec{\sigma}^* \cdot \vec{p} \; \psi_- \right) \label{hmeta}
\end{eqnarray}
where $\psi_\pm \equiv \left(\begin{array}{c} b_\pm \\ a_\pm \end{array}\right)$ are two-component Dirac spinors, and $\vec{\sigma} \equiv (\sigma_1, \sigma_2)$, $\vec{\sigma}^* \equiv (\sigma_1, - \sigma_2)$, with $\sigma_i$ the Pauli matrices. Notice that here the $1/2$-spinor description emerges from the two sublattice honeycomb structure instead of the intrinsic spin of the $\pi$ electron.

Hence, if one considers the linear/relativistic-like regime only, the first scale is
\be
E_\ell \sim v_F / \ell \sim \SI{4.2}{\electronvolt}. \label{Escale}
\ee
Notice that $E_\ell \sim 1.5 \eta$, and that the associated wavelength, $\lambda = 2 \pi / |\vec{p}| \simeq 2 \pi v_F / E$, is $2 \pi \ell$. The electrons' wavelength, at energies below $E_\ell$, is large compared to the lattice length, $\lambda > 2 \pi \ell$. Those electrons see the graphene sheet as a continuum.

The two spinors are connected by the inversion of the full momentum $\vec{k}^D_+ + \vec{p} \to - \vec{k}^D_+ - \vec{p} \equiv \vec{k}^D_- - \vec{p}$. Whether one needs one or both such spinors to describe the physics, strongly depends on the given set-up. For instance, when only strain is present, one Dirac point is enough (see, e.g., \cite{ipfirst}), similarly (see below here) when certain approximations on the curvature are valid~\cite{iorio2012,iorio2014,iorio2015}. The importance and relevance of the two Dirac points for emergent descriptions of scenarios of the high-energy theoretical research, has been discussed at length in \cite{ip3}, where the role of grain boundaries, and the related necessity for two Dirac points, was explained in terms of a relation to spacetime torsion, see below. The full focus on torsion, though, is in \cite{ip4}.

When only one Dirac point is necessary over the whole linear regime, the following Hamiltonian well captures the physics of undeformed (planar and unstrained) graphene
\begin{equation}
    H    = - \ii \, v_F \, \int d^2 x \; \psi^\dagger \vec{\sigma} \cdot \vec{\partial} \, \psi \;, \label{HGrapheneBpaper}
\end{equation}
where the two-component spinor is, e.g., $\psi \equiv \psi_+$, we moved back to configuration space, $\vec{p} \to - \ii \, \vec{\partial}$, and sums turned into integrals because of the continuum limit. In various papers, this regime was exploited to a great extent till the inclusion of curvature and torsion in the geometric background. On the other hand, the regimes beyond the linear one were also investigated. There, granular effects associated with the lattice structure emerge, see~\cite{egypt2018} and also the related~\cite{GUPBTZ}.

When both Dirac points are necessary, one needs to consider four component spinors in a reducible representation \cite{Gusynin,iorio2015,Hands}
$\Psi \equiv \left( \begin{array}{c} \psi_+ \\ \psi_- \\ \end{array} \right)$, and $4 \times 4$ Dirac matrices $\alpha^i = \left(\begin{array}{cc} \sigma^i & 0 \\ 0 & - {\sigma^*}^i \\ \end{array} \right)$, $\beta = \left(\begin{array}{cc} \sigma^3 & 0 \\ 0 & \sigma^3 \\ \end{array} \right)$, $i = 1, 2$. These matrices satisfy all the standard properties, see, e.g., \cite{ip3} and \cite{iorio2015}.

With these, the Hamiltonian is
\begin{equation}
H  =  - \ii \, v_F \int d^2 x \left( \psi_+^\dagger \vec{\sigma} \cdot \vec{\partial} \; \psi_+
    - \psi_-^\dagger \vec{\sigma}^* \cdot \vec{\partial} \; \psi_- \right) =
    - \ii \, v_F \int d^2 x \; \bar{\Psi} \vec{\gamma} \cdot \vec{\partial} \; \Psi \;.  \label{Hdiracgraphene1}
\end{equation}

\subsection{Second scale, $E < E_r < E_\ell$: from the flat space to curved space Dirac Hamiltonian}\label{SubSecSecondScale}

In \cite{i2}, the goal was to identify the conditions for graphene to get as close as possible to a full-power QFT in curved spacetime. Therefore, key issues had to be faced, such as the proper inclusion of the time variable in a relativistic-like description and the role of the nontrivial vacua and their relation to different quantization schemes for different observers. All this finds its synthesis in the Unruh or the Hawking effects, the clearest and unmistakable signatures of QFT in curved spacetime. Therefore, starting from \cite{i2,weylgraphene}, this road was pursued in \cite{iorio2012,iorio2014}. Let us explain here the main issues and the approximations made there.

Besides the scale (\ref{Escale}), when we introduce curvature, we also have a second scale. When this happens, $E_\ell$ is our ``high energy regime'', as we ask the curvature to be small compared to a maximal limiting curvature, $1/\ell^2$, otherwise: i) it would make no sense to consider a smooth metric, and ii) $r < \ell$ (where $1/r^2$ measures the intrinsic curvature), means that we should bend the very strong $\sigma$-bonds, an instance that does not occur. Therefore, our second scale is
\be
E_r \sim v_F / r  \;,
\ee
with $E_r =  \ell / r \; E_\ell  < E_\ell$. To have a quantitative handle on these scales, let us take, e.g., $r \simeq 10 \ell$ as a small radius of curvature (high intrinsic curvature). To this corresponds an energy $E_r \sim \SI{0.4}{\electronvolt}$, whereas to $r \sim \SI{1}{\milli\metre} \sim 10^6\, \ell$, corresponds $E_r \sim \SI{4}{\micro\electronvolt}$. The ``high energy'' to compare with is $E_\ell \sim \SI{4}{\electronvolt}$.

When energies are within $E_r$ (wavelengths comparable to $2 \pi r$), the electrons experience the global effects of curvature. That is to say, at those wavelengths, they can distinguish between a flat and curved surface and, in particular, between, e.g., a sphere and a pseudosphere.  Therefore, whichever curvature $r > \ell$ we consider, the curvature effects are felt until the wavelength becomes comparable to $2 \pi \ell$. The formalism we have used, though, considers all deformations of the geometric kind, except for torsion. Hence, this includes intrinsic curvature and elastic strain of the membrane (on the latter, see \cite{ipfirst}). However, the power stops before $E_\ell$, because there local effects (such as the actual structure of the defects) play a role that must be taken into account in a QG type of theory. On the latter, the first steps were moved in \cite{egypt2018} and also in \cite{GUPBTZ} and in the forthcoming \cite{cagliari}.

The intrinsic curvature is taken here as produced by disclination defects, that are customarily described in elasticity theory (see, e.g., \cite{Kleinert}), by the (smooth) derivative of the (non-continuous) $SO(2)$-valued rotational angle $\partial_i {\omega} \equiv {\omega_i}$, where $i=1,2$ is a ``curved'' spatial index\footnote{Here our notations: $\mu , \nu = 0, 1, ..., n-1$ are Einstein indices, responding to diffeomorphisms, $a, b = 0, 1, ..., n-1$ are flat indices, responding to local Lorentz transformations, while $\alpha,\beta$ are spin indices. The covariant derivative is
\be
\nabla_\mu \psi_\alpha = \partial_\mu \psi_\alpha + {\Omega_\mu}_\alpha^{\; \beta} \psi_\beta \nonumber \;,
\ee
with
\be
{\Omega_\mu}_\alpha^{\; \beta} =  \frac{1}{2} \omega_\mu^{a b} (J_{a b})_\alpha^{\; \beta} \nonumber \;,
\ee
where $(J_{a b})_\alpha^{\; \beta}$ are the Lorentz generators in spinor space, and
\be
{\omega_\mu}^a_{\; b} = e^a_\lambda (\delta^\lambda_\nu \partial_\mu + \Gamma_{\mu \nu}^\lambda) E^\nu_b \nonumber \;,
\ee
is the spin connection, whose relation to the Christoffel connection comes from the full metricity condition $\nabla_\mu e^a_\nu = \partial_\mu e^a_\nu - \Gamma_{\mu \nu}^\lambda e^a_\lambda +  \omega_{\mu \; b}^a e^b_\nu = 0$. We also introduced the Vielbein $e^a_\mu$ (and its inverse $E_a^\mu$), satisfying $\eta_{a b} e^a_\mu e^b_\nu = g_{\mu \nu}$, $e^a_\mu E_a^\nu = \delta_\mu^\nu$, $e^a_\mu E_b^\mu = \delta_b^a$, where $\eta_{a b} = {\rm diag} (1, -1, ...)$. The Weyl dimension of the Dirac field $\psi$ in $n$ dimensions is $d_\psi = (1 - n)/2$. Here $n=3$, and we can move one dimension up (embedding), or down (boundary). More notations can be found in \cite{i2}.}. The corresponding (spatial) Riemann curvature tensor is easily obtained
\begin{equation}\label{a11}
    {R^{i j}}_{k l} =
    \epsilon^{i j} \epsilon_{k l} \epsilon^{m n} \partial_{m} \omega_{n} =
    \epsilon^{i j} \epsilon_{l k} 2 {\cal K}.
\end{equation}
where $\cal K$ is the Gaussian (intrinsic) curvature of the surface. In this approach we have included time, although the metric we adopted is
\begin{equation}\label{mainmetric}
g^{\rm graphene}_{\mu \nu}  = \left(\begin{array}{cc} 1 & 0  \quad 0 \\ \begin{array}{c} 0 \\ 0 \end{array} & - g_{i j} \\ \end{array} \right)\;,
\end{equation}
i.e., the curvature is all in the spatial part, and $\partial_t g_{i j}= 0$. Since the time dimension is included, the SO(2)-valued (abelian) disclination field has to be lifted-up to a SO(1,2)-valued (non-abelian) disclination field\footnote{Recall that in three dimensions $\omega_{\mu \; a b} = \epsilon_{a b c} \,\omega_\mu^{\;\; c}$.}, ${\omega_\mu}^a$, $a=0,1,2$, with $\omega_\mu^{\; a} = e^b_\mu \omega_b^{\; a}$ and the expression
\begin{equation}\label{omega3d}
\omega_a^{\; d}  = \frac{1}{2} \epsilon^{b c d} \left( e_{\mu a} \partial_b E_c^\mu + e_{\mu b} \partial_a E_c^\mu + e_{\mu c} \partial_b E_a^\mu \right) \,,
\end{equation}
gives the relation between the disclination field and the metric (dreibein). All the information about intrinsic curvature does not change. For instance, the Riemann curvature tensor, ${R^\lambda}_{\mu \nu \rho}$, has only one independent component, proportional to $\cal K$, just like in (\ref{a11}) (see \cite{i2}).

With all of the above in mind, the hypothesis is that, when only curvature is important, the long wavelength/small energy electronic properties of graphene, are well described by the following action
\begin{equation}\label{actionAcurvedpaper}
{\cal A} = \ii \, v_F \int d^3 x \sqrt{g} \; \bar{\Psi} \gamma^\mu (\partial_\mu + \Omega_\mu) \Psi \;,
\end{equation}
with $\Omega_\mu \equiv {\omega_\mu}^a J_a$, and $J_a$ are the generators of SO(1,2), the local Lorentz transformations in this lower-dimensional setting. Notice that $J_a$ can never take into account mixing of the $\psi_\pm$,  because they are of the form $J^a = \left(\begin{array}{cc} j^a_+ & 0 \\ 0 & j^a_- \\ \end{array} \right)$, whereas, what is necessary are generators of the form $K^a = \left(\begin{array}{cc} 0 & k^a_+ \\  k^a_- & 0 \\ \end{array} \right)$. This point was discussed at length in \cite{ip3}, within the Witten approach \cite{witten3dgravity}. In that approach, the most general gauge field, that takes into account curvature (intrinsic and extrinsic) and torsion has the following structure $A_\mu = \Omega_\mu + K_\mu$, where $K_\mu \equiv {e^a}_\mu K_a$, hence a Poincar\'{e} ($ISO(2,1)$) or (A)dS type of gauge theory, depending on the role played in here by the cosmological constant (on this see \cite{iorio2012,iorio2014}, and the review \cite{iorio2015}). The matter, though, might be faced by taking an alternative view, for which the gauge fields are internal rather than spatiotemporal. In this case, a link with the supersymmetry (SUSY) introduced in \cite{AVZ} (that is a SUSY without superpartners, often referred to as \textit{unconventional} SUSY (USUSY)) can be established, as is shown in \cite{ip3} and in \cite{APZ,DauriaZanelli2019,ADVZ2021}, as is briefly discussed in Section \ref{GB_section}.

Let us clarify here an important point. Within this scenario, a nontrivial $g_{00}$ in (\ref{mainmetric}), hence a clean nontrivial general relativistic effect (recall that $g_{00} \sim V_{\mbox{grav}}$) can only happen if specific symmetries and set-ups map the lab system into the wanted one. A lot of work went into it, e.g., \cite{iorio2012,iorio2014}, and went as far as producing measurable predictions of a Hawking/Unruh effect, for certain specific shapes. Let us recall here the main ideas behind this approach, which we may call the ``Weyl symmetry approach'' \cite{iorio2015}.

\subsection{The importance of Weyl symmetry}\label{SubSecWeylSymm}

First of all, one notices that the action (\ref{actionAcurvedpaper}) enjoys local Weyl symmetry
\begin{equation}
g_{\mu \nu} \to e^{2 \sigma(x)} g_{\mu \nu} \, \quad {\rm and} \quad \Psi \to e^{-\sigma(x)} \Psi \;,
\end{equation}
that is an enormous symmetry among fields/spacetimes \cite{lor}. As explained in \cite{i2,weylgraphene}, to make the most of the Weyl symmetry of (\ref{actionAcurvedpaper}), we better focus on conformally flat metrics. The simplest metric to obtain in a laboratory is of the kind (\ref{mainmetric}). For this metric the Ricci tensor is ${R_\mu}^\nu = {\rm diag}(0, {\cal K}, {\cal K})$. This gives as the only nonzero components of the Cotton tensor, $C^{\mu \nu} = \left( \epsilon^{\mu \sigma \rho} \nabla_\sigma {R_\rho}^\nu + \mu \leftrightarrow \nu\right)$, the result $C^{0 x} = - \partial_y {\cal K} = C^{x 0} $ and $C^{0 y} = \partial_x {\cal K} = C^{y 0}$. Since conformal flatness in $(2+1)$ dimensions amounts to $C^{\mu \nu} = 0$, this shows that all surfaces of constant $\cal K$ give raise in (\ref{mainmetric}) to conformally flat $(2+1)$-dimensional spacetimes. This points the light-spot to surfaces of constant Gaussian curvature.

The result $C^{\mu \nu} = 0$ is intrinsic (it is a tensorial equation, true in any frame), but to exploit Weyl symmetry to extract non-perturbative exact results, we need to find the coordinate frame, say it $Q^\mu \equiv (T,X,Y)$, where
\begin{equation}\label{genexplicitconfflat}
g^{\rm graphene}_{\mu \nu}  (Q) = \phi^2(Q) g^{\rm flat}_{\mu \nu} (Q) \;.
\end{equation}
Besides the technical problem of finding these coordinates, the issue to solve is the physical meaning of the coordinates $Q^\mu$, and their practical feasibility. See \cite{iorio2015}, and \cite{ioriokus}.

Tightly related to the previous point is the conformal factor that makes the model {\it globally predictive, over the whole surface/spacetime}. The simplest possible solution would be a single-valued, and time independent $\phi(q)$, already in the original coordinates frame, $q^\mu \equiv (t,u,v)$, where $t$ is the laboratory time, and, e.g., $u, v$ the meridian and parallel coordinates of the surface.

Here we are dealing with a spacetime that is embedded into the flat $(3+1)$-dimensional Minkowski. Although, as said, the focus is on intrinsic curvature effects, just like in a general relativistic context, issues related to the embedding, even just for the spatial part, are important. For instance, when the surface has negative curvature, one needs to move from the abstract objects of non-Euclidean geometry, to objects measurable in a Euclidean real laboratory. This involves the last point above about global predictability, and, in the case of negative curvature, necessarily leads to singular boundaries for the surfaces, as proved in a theorem by Hilbert, see, e.g., \cite{iorio2015} and \cite{icrystals}. Even the latter fact is, once more, a coordinates effect, due to our insisting in embedding a negative curvature surface in $\mathbb{R}^3$, and clarifies the hybrid nature of these emergent relativistic settings. The quantum vacuum of the field that properly takes into account the measurements processes, as for any QFT on a curved spacetime, was identified, including how the graphene hybrid situation can realize that \cite{iorio2012,iorio2014}. As well known, this is crucial in QFT, in general, and on curved space, in particular.

The above lead us to propose a variety of set-ups, the most promising being the one obtained by shaping graphene as a Beltrami pseudosphere \cite{iorio2012,iorio2014,iorio2015}, a configuration that can be put into contact with three key spacetimes with horizon: the Rindler, the de Sitter and the Ba\~{n}ados-Teitelboim-Zanelli (BTZ) BH \cite{BTZ1992}. The predicted impact on measurable quantities is reported in the first papers, and then explored in the subsequent efforts of computer-based simulations.

\subsection{Ramifications}\label{SubSecRamifications}

Many other high energy scenarios can be reached with graphene and related systems that go under the name of Dirac materials (DMs) \cite{wehling}. Here we list some of such directions.

\subsubsection{Generalized Uncertainty Principles on DMs}

In \cite{egypt2018} (see also \cite{GUPBTZ}), the realization in DMs of specific generalized uncertainty principles (GUPs) associated with the existence of a fundamental length scale was studied. The scenarios that one wants to reproduce there is that for which the commutation relations are modified, by quantum gravity effects, to be (see, e.g., \cite{Maggiore:1993rv,Kempf:1993bq,Scardigli:1999jh,alidasvagenas1,alidasvagenas2,alidasvagenas3,Buoninfante:2019fwr,Petruzziello:2020wkd,Bosso:2021koi,Naveed2022} and references therein)
\be\label{ADV comm rel}
[x_i,p_j] = \ii  \, \hbar \, \left( \delta_{i j} - A \left( |\vec{p}| \delta_{i j} + \frac{p_i \, p_j}{|\vec{p}|} \right) + A^2 \left( |\vec{p}|^2 \delta_{i j} + 3 p_i p_j  \right) \right) \,,
\ee
where $A = \tilde{A} \, \ell_P / \hbar$, with $\tilde{A}$ a phenomenological dimensionless parameter and $\ell_P \sim \SI{e-35}{\metre}$ the Planck length.

In \cite{egypt2018}, it is shown that a generalized Dirac structure survives beyond the linear regime of the low-energy dispersion relations. Also, a GUP of the kind compatible with (\ref{ADV comm rel}), related to QG scenarios with a fundamental minimal length (there the graphene lattice spacing) and Lorentz violation (there the particle/hole asymmetry, the trigonal warping, etc.) is naturally obtained. It is then shown that the corresponding emergent field theory is a table-top realization of such scenarios by explicitly computing the third-order Hamiltonian and giving the general recipe for any order. Remarkably, these results imply that going beyond the low-energy approximation does not spoil the well-known correspondence with analog massless quantum electrodynamics phenomena (as usually believed). Instead, it is a way to obtain experimental signatures of quantum-gravity-like corrections to such phenomena.

In \cite{GUPBTZ}, the authors investigated the structure of the gravity-induced GUP in $(2+1)$-dimensions. They showed that the event horizon of the $M \neq 0$ BTZ micro-black-hole furnishes the most consistent limiting ``gravitational radius'' $R_g$ (that is, the fundamental minimal length induced by gravitational effects). A suitable formula for the GUP and estimate the corrections induced by the latter on the Hawking temperature and Bekenstein entropy could be obtained. As for the role of graphene, it is shown that the extremal $M = 0$ case, and its natural unit of length introduced by the cosmological constant, $\ell = 1/\sqrt{- \Lambda}$, is a possible alternative to $R_g$, and DMs, when shaped as hyperbolic pseudospheres, represent condensed matter analog realizations of this scenario with $\ell = \ell_{DM}$. Due to the peculiarities of three-dimensional gravity \cite{DJtH}, this configuration can still be regarded as a BH, even though $M = 0$, on this see, e.g., \cite{BHTZ,modave,wittenM=0}.

More work in this QG phenomenology direction is forthcoming \cite{cagliari}. There it is found that even more GUPs are at work at different energy scales, and a link is established between the abstract coordinates satisfying the GUPs and the coordinates one measures in the lab.

With this in mind, one sees that our scales here are much more within reach than those of (\ref{ADV comm rel}). Indeed $\ell_P$ needs to be traded for the lattice spacing $\ell$, that, e.g., for graphene is $\ell_{graphene} \sim \SI{1.4e-10}{\metre}$. Therefore, we have much more hope to see in DMs the effects of the modifications to $[x_i,p_j] = \ii \, \hbar \, \delta_{i j}$ compared with the direct effects of $O(\ell_P)$.

\subsubsection{Grain boundaries on DMs and two scenarios: Witten $3D$ gravity, and USUSY}
\label{GB_section}

In \cite{ip3}, two different high-energy-theory correspondences on DMs associated with {\it grain boundaries} (GBs) are proposed. We recall here that a GB can be realized as a line of disclinations of opposite curvature, for instance, pentagons and heptagons, arranged so that two regions (grains) of the membrane match. These grains have different relative orientations, given by the so-called misorientation angle $\theta$, which characterizes the GB defect. Each side of the GB corresponds to one of the Dirac points (and the other is related by a parity transformation, see Appendix B of \cite{ip3} for details) in the continuous $\pi$ electron description. Therefore, the continuous limit description of the $\pi$ electrons living in a honeycomb with GB needs the two inequivalent Dirac points. Even more, as the $\theta$ angle is related to a non-zero Burgers vector $\vec{b}$ through the Frank formula, and a non-zero $\vec{b}$ implies non-zero torsion in the continuous limit\footnote{Roughly speaking, torsion is the surface density of the Burgers vector $\vec{b}$. For technicalities, see \cite{Kleinert,Katanaev1992}.}, such description should take into account torsion.

The first correspondence points to a $(3+1)$-dimensional theory, with spatiotemporal gauge group $SO(3,1)$, with nonzero torsion, locally isomorphic to the Lorentz group in $(3+1)$ dimensions, or the de Sitter group in $(2+1)$ dimensions, in the spirit of $(2+1)$-dimensional gravity \`{a}~la  Witten~\cite{witten3dgravity}. The other correspondence treats the two Dirac fields as an internal symmetry doublet, and it is linked there with USUSY~\cite{AVZ} with $SU(2)$ internal symmetry \cite{APZ}. One of the properties of USUSY is the absence of gravitini, although it includes gravity and supersymmetry. Even if in $(2+1)$ dimensions it is constructed from a Chern-Simons connection containing fermion fields, the only propagating local degrees of freedom are the fermions \cite{GPZ}. Notice that in USUSY, the torsion of geometric backgrounds appears naturally, and its totally-antisymmetric part is coupled with fermions.

Those results pave the way for the inclusion of GB in the emergent field theory picture associated with these materials, whereas disclinations and dislocations have already been well explored.

\subsubsection{Particle-Hole pairs in graphene to spot spatiotemporal torsion}

In \cite{ip4}, assuming that dislocations could be meaningfully described by torsion, a scenario is proposed based on the role of time in the low-energy regime of two-dimensional DMs, for which coupling of the fully antisymmetric component of the torsion with the emergent spinor is not necessarily zero. That approach is based on the realization of an exotic \textit{time-loop}, that could be seen as oscillating particle-hole pairs. Although that is a theoretical paper, the first steps were moved toward testing the laboratory realization of these scenarios by envisaging \textit{Gedankenexperiments} on the interplay between an external electromagnetic field (to excite the pair particle-hole and realize the time-loops) and a suitable distribution of dislocations described as torsion (responsible for the measurable holonomy in the time-loop, hence a current). The general analysis establishes that we need to move to a nonlinear response regime. Then the authors conclude by pointing to recent results from the interaction laser-graphene that could be used to look for manifestations of the torsion-induced holonomy of the time-loop, e.g., as specific patterns of suppression/generation of higher harmonics. As said before, USUSY takes into account torsion and couples its totally antisymmetric component with fermions in a very natural way. Therefore, it could play a significant role also in this exotic time loop \cite{ichep2020}.


\subsubsection{Vortex solutions of Liouville equation and quasi spherical surfaces}

In \cite{ioriokus}, the authors identified the two-dimensional surfaces corresponding to specific solutions of the Liouville equation of importance for mathematical physics, the non-topological Chern-Simons (or Jackiw-Pi \cite{jackiwpi1,jackiwpi2}) vortex solutions, characterized by an integer \cite{Horvathy_Yera} $N \ge 1$. Such surfaces, called there $S^2 (N)$, have positive constant Gaussian curvature, $K$, but are spheres only when $N=1$. They have edges and, for any fixed $K$, have maximal radius $c$ that is found there to be $c = N / \sqrt{K} $. If such surfaces are constructed in a laboratory using DMs, these findings could be of interest to realize table-top Dirac massless excitations on nontrivial backgrounds. Then the types of three-dimensional spacetimes obtained as the product $S^2 (N) \times \mathbb{R}$ are  also briefly discussed.

\subsection{Realization in the labs}\label{SubSecExperiment}

Besides the theoretical work just outlined, one should always aim at the actual realization of the necessary structures in real laboratories. See, e.g., the work~\cite{icrystals}, where Lobachevsky geometry was realized via simulations by producing a carbon-based mechanically stable molecular structure arranged in the shape of a Beltrami pseudosphere. It was found there that this structure: i) corresponds to a non-Euclidean crystallographic group, namely a loxodromic subgroup of $SL(2,\mathbb{Z})$; ii) has an unavoidable singular boundary that is fully taken into account. That approach, substantiated by extensive numerical simulations of Beltrami pseudospheres of different sizes, might be applied to other surfaces of constant negative Gaussian curvature and points to a general procedure to generate them. Such results pave the way for future experiments. More work is currently undergoing.

\section{Graphene-inspired quantum gravity: the quasiparticle picture}
\label{SecBekensteinXons}

If the entropy of any physical system of volume $V$, including the entropy associated to space itself, is never bigger than the entropy of the BH whose event horizon coincides with the boundary of $V$ \cite{bekensteinBound1}
\begin{equation}\label{eq:BekensteinBound}
	S \leq S_{BH} \,,
\end{equation}
this means that the associated Hilbert space, $\HH$, has finite dimension, ${\rm dim} (\HH) \sim e^{S_{BH}}$. This simple consideration poses serious questions.

In fact, at our energy scales, the world is well described by fields (matter) and the space they live in. Quantum fields, as we know them, act on infinite-dimensional Hilbert spaces, to which one should add the degrees of freedom surely carried by (the quanta of) space itself. How can then be that the {\it ultimate} Hilbert space, which must include all degrees of freedom, is not only separable, like for a single harmonic oscillator, but it is actually finite-dimensional?

This logic points to the existence of something more fundamental, making both matter and space. Hence, the \textit{elementary} particles of the Standard Model (leptons, quarks, etc) would be, in fact, quantum \textit{quasi-particles}, whose physical properties (spin, mass, etc) are the effect of the interaction with a \textit{lattice} whose emergent picture is, in turn, (classical) space. Inspired by Feynman \cite{Feynman} (see the Introduction here) these objects were called $X$\textit{ons} \cite{Xons}. To access the $X$ons one needs resolutions of the order of the Planck length, which might not only be technically unfeasible, but actually impossible, see, e.g., \cite{Doplicher1995}.

In~\cite{aischol}, and later in~\cite{aismal}, general arguments are provided regarding the connection between our low-energy quantum-matter-on-classical-space description and an hypothetical fundamental theory of the $X$ons. The reshuffling of the fundamental degrees of freedom during the unitary evolution then leads to an entanglement between space and matter. The consequences of such scenario are considered in the context of BH evaporation, see, e.g., \cite{page2,page1,harlow}, and the related information loss: a simple toy model is provided in which an average loss of information is obtained as a consequence of the entanglement between matter and space. Pivotal for the previous study is the work of~\cite{alfgae}, where the Hawking--Unruh phenomenon is studied within an entropy-operator approach, \textit{\`{a}~la} Thermo-Field Dynamics (TFD)~\cite{TFD1,TFD2} that discloses the thermal properties of BHs.

\subsection{The universal quasiparticle picture}\label{SubQuasiPartPict}

Emergent, nonequivalent descriptions of the same underlying dynamics are ubiquitous in QFT~\cite{Dirac1966}, as, in general, the vacuum has a nontrivial structure with nonequivalent\footnote{Let us explain why we use the word \textit{phase} in quotation marks. Given the general vacuum of a QFT, one can identify several vacua that cannot be obtained one from the other through a smooth unitary transformation. Starting from each of these ``sub vacua'', and acting with the appropriate creation operators, one builds several (infinite) sectors, sometimes called super-selection sectors. Not all of them correspond to a phase of the system, in the proper statistical mechanical/thermodynamical sense. On the other hand, all such phases need be described by a super-selection sector or by a set of them. On this, see, e.g., \cite{umezawa1993advanced}.} ``phases'' \cite{TFD2}. That is, for a given basic dynamics (governed by an Hamiltonian or a Lagrangian) one should expect several different Hilbert spaces, representing different ``phases'' of the system with distinct physical properties. Distinct excitations play the role of the elementary excitations for the given ``phase'' but their general character is that of the quasiparticles of condensed matter~\cite{TFD1,TFD2}.

What it is added here to that QFT picture is that
\begin{itemize}
 \item the degrees of freedom are finite, hence fields are necessarily emergent;
 \item spacetime is also emergent.
\end{itemize}
Taking this view, the continuum of fields and space is then only the result of an approximation, of a limiting process. In general, there must be (many!) microscopic configurations of the $X$ons giving rise to the \textit{same} emergent space but to \textit{different/non-equivalent} fields.

With this in mind (for details see \cite{aischol}), the generic state $\ket{\psi}\in\HH$ can be written as
\begin{equation} \label{efbhstate}
  \ket{\psi} = \bigoplus_{i=1}^{N_T} \sum_{I=1}^{p_i} \sum_{n=0}^{q_i-1} c^{(i)}_{In} \ket{I_i} \otimes \ket{n_i},
\end{equation}
where the vectors $\ket{I_i}$ and $\ket{n_i}$ form a basis of $\HH_{\mathrm{G}}^{p_i}$ and $\HH_{\mathrm{F}}^{q_i}$, that are the Hilbert space of the ``spatial degrees of freedom'' (Geometry) of dimension $p_i$ and of the Hilbert space of the ``matter degrees of freedom'' (Fields) of dimension $q_i$, respectively, and $c^{(i)}_{In}$ are numerical coefficients. Notice that $N_T$ is the number of specific rearrangements (Topologies) of the degrees of freedom.

By denoting with ${\PP}_{(i)} : \HH \mapsto T_{(i)}$ a projector onto $T_{(i)}$, a subspace with a given ``Topology'', the associated density matrix, representing the state of the field, is
\begin{equation}\label{eq:rho-i}
  {\rho}_{(i)} = \Tr_{\HH_{\mathrm{G}}^{p_i}} \ket{\psi}_{i} \bra{\psi}_i ,
\end{equation}
where $\ket{\psi}_i \sim {\PP}_{(i)}\ket{\psi}$, and we trace away the degrees of freedom of the gravitational field. Correspondingly, the entropy of entanglement between matter and space, for a given topology of the lattice, is the usual expression\footnote{As it is impossible to distinguish the space corresponding to different topologies of the lattice, the expected value of the entanglement between the fields and the geometrical degrees of freedom is $\langle S \rangle = \sum_i p_{(i)} S_{(i)}$.}
\begin{equation}\label{eq:S-i}
  S_{(i)} = - \Tr_{\HH_{\mathrm{F}}^{q_i}} {\rho}_{(i)} \ln {\rho}_{(i)} \,.
\end{equation}

This picture needs to be compared to the standard QFT picture, recalled earlier, of the non-equivalent field configurations, or ``phases'' \textit{\`{a}~la} TFD~\cite{TFD1,TFD2} where the mirror degrees of freedom, that characterize TFD (often called there the \textit{tilde} degrees of freedom), model the degrees of freedom of the geometry. These degrees of freedom are then traced away, leaving us with quantities all referring solely to matter (fields). Indeed, the vacuum of TFD can be written as~\cite{TFD1}
\begin{equation}\label{eq:TFDvacuum}
	\ket{0 (\theta)} = \sum_n \sqrt{w_n(\theta)} \ket{n, \tilde{n}} ,
\end{equation}
where $\theta$ is a physical parameter labeling the different ```phases'', $w_n$ are probabilities such that $\sum_n w_n =1$, and the states $\ket{n, \tilde{n}}$ (infinite in number) are the components of the condensate, each made of pairs of $n$ quanta and their $n$ mirror counterparts ($\tilde{n}$). Therefore, such vacuum is clearly an entangled state. Notice that~\cite{TFD1}
\begin{equation}\label{eq:TFDinequiv}
	\bra{0(\theta)} 0 (\theta') \rangle \to 0 ,
\end{equation}
in the field limit, which formalizes the inequivalence we have discussed. Notice also that, if one fixes $\theta$, there is no unitary evolution to disentangle the vacuum, as the interaction with the environment and non-unitarity are the basis for the generation and the stability of such entanglement \cite{alfgae}.

The expected value of \textit{field's} observables, $O$, are obtained by tracing away the \textit{mirror} modes, $\tilde{n}$. In the TFD formalism this corresponds to taking the vacuum expectation value over the vacuum (\ref{eq:TFDvacuum})
\begin{equation}\label{eq:TFDOavg}
\langle O \rangle \equiv \langle 0(\theta) | O | 0 (\theta) \rangle = \sum_n \, w_n(\theta) \, \langle n | O | n \rangle  \,.
\end{equation}

In particular, there is always an \textit{entanglement entropy} associated to any field, given by, e.g.,
\begin{equation}\label{eq:TFDentropy}
\langle S \rangle = \sum_k \left[ n_k \ln n_k + (1 - n_k) \ln (1 - n_k) \right] \,.
\end{equation}
where $n_k = \langle N_k \rangle$ is the expected value of the number operator for the given (fermionic, in this example) mode $k$. The analogy of (\ref{eq:TFDentropy}) with (\ref{eq:S-i}) is stronger, if we think that in TFD the process of taking statistical averages through tracing is replaced, by construction~\cite{TFD1}, by taking vacuum expectation values (vevs) over the vacuum (\ref{eq:TFDvacuum}). Furthermore, as well known, in the basis where the density matrix in the entropy (\ref{eq:S-i}) is diagonal, the entropy can be written as
\begin{equation}\label{eq:TFDentropyWn}
\langle S \rangle = - \sum_n w_n(\theta) \ln w_n(\theta) \,,
\end{equation}
as shown, e.g., in~\cite{TFD2}.

In this comparison, the mirror (tilde) image of the field mimics the effects of the entanglement with space where the field lives, even when the space is flat. This happens on a level that is both emergent and effective. This would have far reaching consequences, surely worth a serious exploration. For instance, the entanglement entropy associated to any field, would never be zero. Furthermore, this would explain why the attempt to quantize gravity as we quantize the matter fields, cannot make much sense.

To compare TFD entropies and the entropies obtained in the quasi-particle picture, a different point of view is taken in~\cite{aismal}. There the authors focus on BH evaporation as seen form the point of view of the fundamental $X$ons, and were able to establish formulae and structures indeed similar to those of TFD. The main difference with TFD is that, at the level of the discrete structures related to $X$ons, the quantum field theoretical considerations illustrated above are only an approximation. In Subsection \ref{SubBHevaporationII} we shall recall those results. Before doing so, let us focus on BH evaporation as seen from the point of view of the emergent quantum fields and emergent space.

\subsection{Effects of the quasiparticle picture on black hole evaporation}\label{SubBHEvaporationI}

When applied to $\BH$ evaporation, the immediate consequence of the above is that it is impossible that after the evaporation we can retrieve the very same ``phase''  we had before the $\BH$ was formed. Hence, the information associated to the quantum fields before the formation of the $\BH$ is, in general, lost after the $\BH$ has evaporated, due to the entanglement between matter and space.

Even when the emergent spaces, before the formation and after the evaporation, are the same (say they are both Minkowski spacetimes), the emergent fields belong, in general, to non-equivalent Hilbert spaces. Therefore, even assuming unitary evolution at the $X$ level, the initial and final Hilbert spaces of fields cannot be the same. There is always a \textit{relic} matter-space entanglement entropy.

Looking at Eq. \eqref{efbhstate}, it is clear that the Hilbert space $H$ can be written as
\begin{equation}
\HH = \bigoplus_{i=1}^{N_T} \HH_G^{p_i} \otimes \HH_F^{q_i}   \label{eq:hilbert-space}
\end{equation}
where we can now introduce measures, $R_F$s, $R_G$s, of the ``degenaracies'', $p_i = N_G\,R_G^i$, with $N_G$ classical geometries available (they represent the BH with mass $M^{(a)} = a\,\varepsilon$, where $a=0,1, \dots N_G-1$), and each classical geometry can be realized by $R_G^i$ microstates. On the other hand, $q_i = N_F\,R_F^i$, that is, each emergent field state can be realized by $R_F^i$ indistinguishable microstates.

\begin{figure}
  \centering
  \includegraphics[width=\textwidth]{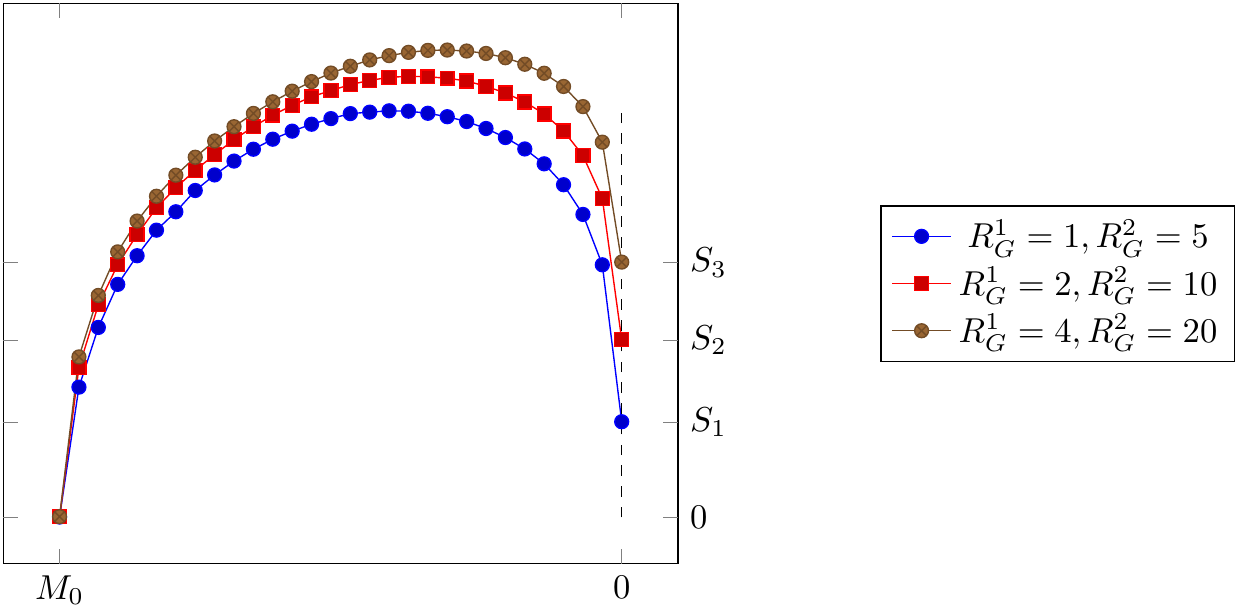}
  \caption{Entropy of the entanglement between matter and space, as a function of the decreasing mass of the evaporating $\BH$. The initial and final points of this curve are in exact correspondence with the initial and final points of the Page curve. The plot here is for two topologies and three cases. The more microscopic realizations of macroscopic classical geometries are allowed, the higher the residual entropies. Here: $S_1 = 0.77$, $S_2 = 1.43$, $S_3 = 2.06$. Figure taken from \cite{aischol}.}
  \label{fig:page-modified_2}
\end{figure}

The analytic computations of the entanglement entropy demand a heavy toll, so in~\cite{aischol} the authors proceeded numerically. The case we present here is for the following choice of $N_G = 30$, $N_T=2$, and $R_F^i=1$, for each topology. The plots in Fig.~\ref{fig:page-modified_2} show the entanglement entropies, corresponding to the three sets of values given in the box, as functions of the discrete evolution parameter $k$.

As can be seen from the figure, the residual entropies are never zero, and are given by
\begin{equation}\label{relicS}
  S_1 = 0.77, \quad S_2 = 1.43, \quad S_3 = 2.06 \,,
\end{equation}
corresponding to the set of values in the box going from the top to the bottom, respectively. The more microscopic realizations of the same macroscopic geometry (i.e. the bigger the degeneracy $R_G$), the higher the relic entanglement entropy. This is as it must be.

\begin{figure}
  \centering
  \includegraphics[width=0.8\textwidth]{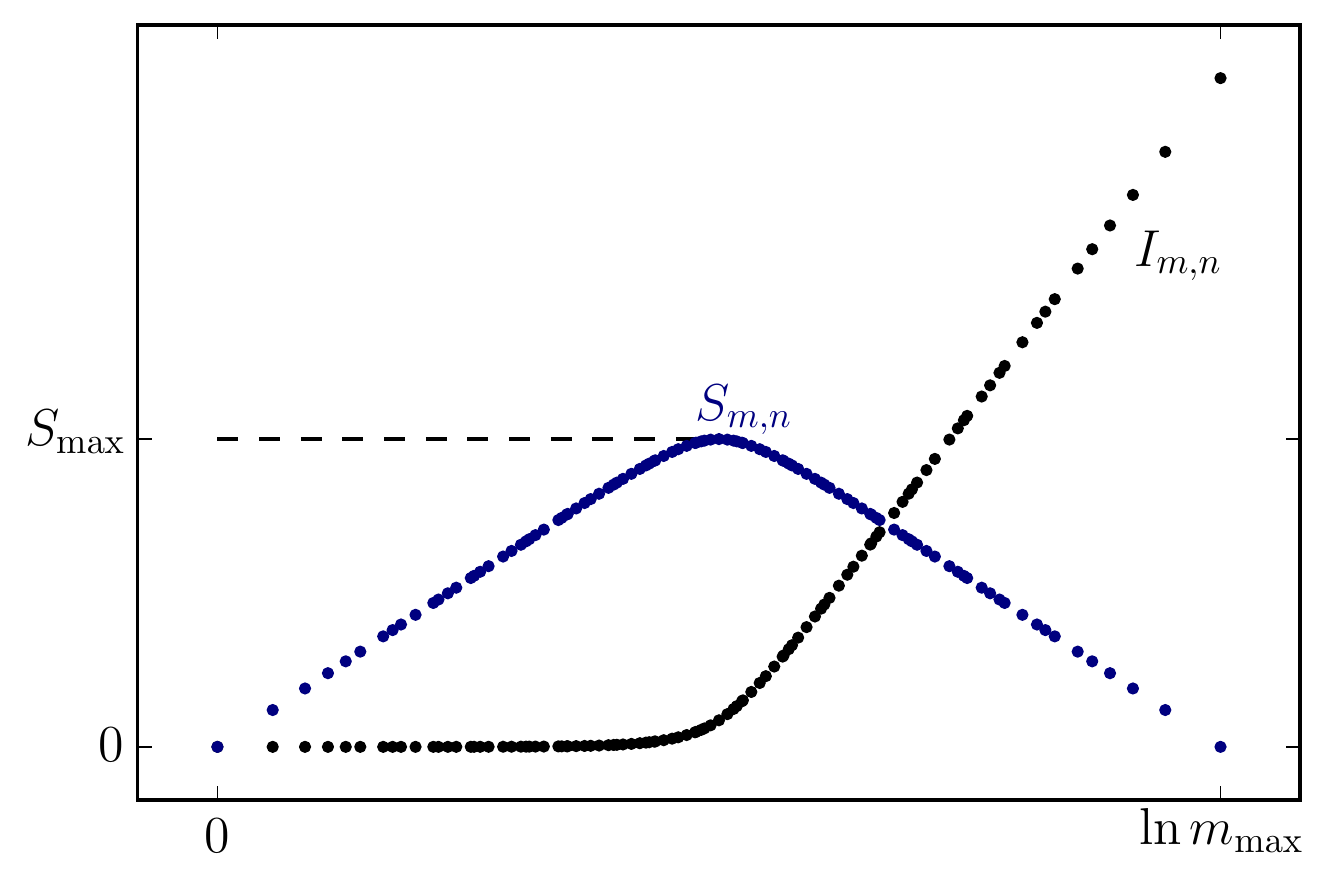}
  \caption{Page curve, representing the entanglement between matter modes inside the $\BH$ and matter modes of the radiation leaving the $\BH$ (in this picture there is no explicit reference to the degrees of freedom related to space) vs the log of the dimension $m$ of the Hilbert space of the radiation, obtained in \cite{page2,page1}. The point $\ln m=0$ corresponds to the initial mass of the $\BH$, $M=M_0$. Indeed, $m=1$ means that only the vacuum state populates the radiation subsystem of the Hilbert space, at the start of the evaporation. On the other hand, $m_{max}$ corresponds a fully evaporated $\BH$, $M=0$. Figure taken from \cite{aischol}.}
  \label{fig:page-original}
\end{figure}

The fact that, at the end of the evaporation, the entanglement entropy remains finite signals a dramatic departure from the {\it information conservation} scenario of the famous Page curve~\cite{page1}, presented here in Fig. ~\ref{fig:page-original}. There the total Hilbert space has dimension $m n$, and consists of two subsystems: the $\BH$ subsystem, of dimension $n \sim e^{A/4}$, where $A$ is the area of the event horizon, and the radiation subsystem, of dimension $m \sim e^{s_{th}}$, where ${s_{th}}$ is the thermodynamic radiation entropy. In Page's picture there is no explicit mention of the degrees of freedom of space and the evolution is taken to be unitary. Thus, in that picture one sees that, when the BH is formed, there is no Hawking radiation outside, hence, $m = 1$ and $n = \dim \HH$. The BH-radiation entanglement entropy, $S_{m,n}$ is trivially zero. As the BH evaporates, $m$ increases, while $n$ decreases, keeping $m \,n$ constant. Since the emitted photons are entangled with the particles under the horizon, $S_{m,n}$ increases, but only up to, approximately, half of the evaporation process. There, the information stored below the horizon starts to leak from the $\BH$, so that $S_{m,n}$ decreases until full evaporation, hence $n = 1$ and $m = \dim \HH$ and $S_{m,n}$ returns to zero.

From the point of view of the quasiparticle picture, we may say that, even if one takes a conservative view for which the $X$ons evolve unitarily, nonunitarity is unavoidable:
\begin{itemize}
  \item the unitary evolution may as well be only formally possible, but physically impossible to measure, for some form of a generalized uncertainty forbidding the necessary Planck scale localization/resolution (see, e.g., \cite{Doplicher1995});
  \item the emergent description of the evolution is that of the combined system gravity+matter, hence there is inevitably information loss, due to the relic entanglement of the matter field with the space;
  \item this description should apply also to standard nonunitary features of QFT, and we evoke here the possibility that the tilde degrees of freedom of TFD could be interpreted as ``how the emergent fields see the degrees of freedom of space with which they are entangled''.
\end{itemize}
Notice that this description does allow for an arbitrary number of different fields, hence naturally includes the possibility of yet unknown (``dark'') kinds of matter.

\subsection{BH evaporation as seen from the $X$ons and the unification of the entropies}\label{SubBHevaporationII}

In \cite{aismal} the authors describe BH evaporation from the point of view of the fundamental constituents, assuming they are fermions, so that only one excitation per quantum level is permitted. Because the $X$ons must be responsible for the formation of both matter and space, no geometric notions can be used. For example, it is assumed that only a finite number $N$ of quantum states/slots are available to the system. This last condition is a non-geometric way of requiring that the system is localized in space. Moreover, it is not meaningful to refer to \emph{interior} and the \emph{exterior} of a BH. Instead the authors there distinguish between \emph{free} and \emph{interacting} $X$ons, respectively: BH evaporation is the process in which the number of \textit{free} $X$ons decreases, $N \to (N -1) \to (N - 2) \to \cdots$, while interacting $X$ons form matter (quasi-particles) and space (geometry), i.e. the environment.

The Hilbert space of \textit{physical states} $H$ is the subspace of a larger \textit{kinematical} Hilbert space $K \ \equiv \ H_\I \otimes H_\II$, and it has dimension $\Sigma \ \equiv \ {\rm dim} \, H \ = \ 2^N$. Here ${\rm I}$ and ${\rm II}$ refer to BH and environment, respectively, in the sense explained above.

The state of the system $|\Psi(\si)\rangle \in H$ is \cite{aismal}
\be \label{alexp}
|\Psi(\sigma)\rangle \ = \ \prod^N_{i=1} \, \sum_{n_i=0,1} \, C_i(\sigma) \, \left(a^{\dag}_i\right)^{n_i} \, \left(b^{\dag}_i \right)^{1-n_i}  \, |0\rangle_{\I} \otimes |0\rangle_{\II} \, ,
\ee
where $a$ and $b$ are environment and BH ladder operators, respectively, and
\be
C_i (\sigma) = (\sin \sigma)^{n_i} \, (\cos \sigma)^{1-n_i} \,.
\ee
$\sigma$ is an interpolating parameter going from $0$ to $\pi/2$. We can also define TFD-like entropy operators
\bea
S_{\I}(\sigma) \ &= \ -\sum^N_{n=1} \, \left(a^\dag_n \, a_n \, \ln \sin^2 \sigma + a_n \, a^\dag_n \, \ln \cos^2 \sigma \right) \, .\\
S_{\II}(\sigma) \ &= \ -\sum^N_{n=1} \, \left(b^\dag_n \, b_n \, \ln \cos^2 \sigma + b_n \, b^\dag_n \, \ln \sin^2 \sigma \right) \, .
\eea
so that their averages on $|\Psi(\sigma)\rangle$ give the von-Neumann entropy of the two subsystems:
\be \label{vnenIII}
\mathcal{S}_\I (\sigma) \ = \ \mathcal{S}_\II  (\sigma) \ = \  - N \left(\sin^2 \sigma \, \ln \sin^2 \sigma + \cos^2\sigma \, \ln \cos^2 \sigma\right) \, .
\ee
Such entropy quantifies the entanglement between the environment and the BH. As for the original Page result, the entropy (\ref{vnenIII}) shows that the BH evaporation at such fundamental level is a unitary process, with $\mathcal{S}(0)=\mathcal{S}(\pi/2)=0$ and a maximum  value $\mathcal{S}_{max} \ = \ N \, \ln 2 \ = \ \ln \Sigma$, so that $\Sigma \ = \  e^{\mathcal{S}_{max}}$. $\mathcal{S}_{max}$ quantifies the maximum information necessary in order to describe the BH and should be identified with the BH entropy before the evaporation. When the BH evaporates the mean number
$ N_\II (\sigma) =  N \, \cos^2 \sigma$ of free $X$ons decreases, while the mean number of interacting $X$ons $ N_\I (\sigma) =  N \, \sin^2 \sigma$ increases. Then, BH and environment entropy should be
\bea
{\cal S}_{BH} \ = \  N \ln 2 \cos^2 \sigma \,, \qquad {\cal S}_{env} \ = \  N \ln 2 \sin^2 \sigma \, .
\eea
Moreover, $\sigma$ finds a natural explanation as a discrete parameter in the interval $[0, \pi/2]$, essentially counting the diminishing number of free $X$ons.

\begin{figure}
\centering
\includegraphics[width=0.70\textwidth]{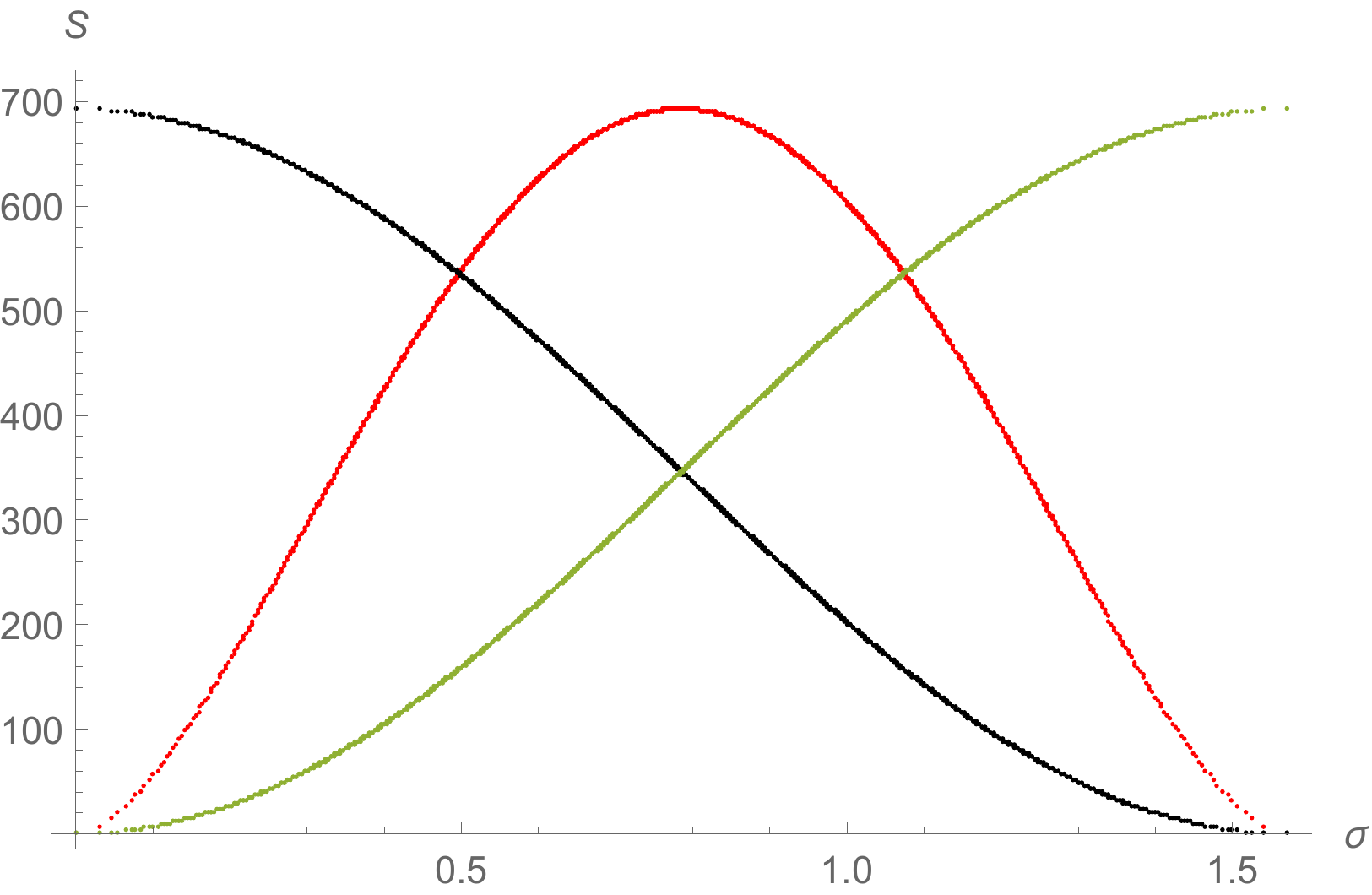}
\caption{Plot of ${\cal S}_{BH}$ (black), ${\cal S}_{env}$ (green) and ${\cal S}_{I}$ (red), as function of $\sigma$, for $N=1000$. Figure taken from \cite{aismal}.}
\label{figen}
\end{figure}

The entanglment, BH and environment entropy satisfy $\mathcal{S}_\I \ \leq \ \mathcal{S}_{BH} \, + \, \mathcal{S}_{env} \ = \  \mathcal{S}_{max}$: the entropy of both BH and environment is bounded from above, in accordance with the Bekenstein bound. In figure \ref{figen}, these three entropies are plotted as functions of the discrete parameter $\si$.
For a full identification of ${\cal S}_{max}$ with the entropy ${\cal S}_{BH}$ of the initial BH, one would have (for a non-rotating, uncharged black hole)
\be
N \ = \ \frac{4 \, \pi \, M^2_0}{l_P^2\, \ln 2} \,,
\ee
where $M_0$ is the initial BH mass. Finally, identifying the $N$ quantum levels with quanta of area, in the spirit of Refs. \cite{mukhanov,bekenstein2,bekenstein3,progress}, one gets
\be
A \ = \ 4 \, N \, l_P^2 \, \ln 2 \, ,
\ee
for the BH horizon area. Notice that the value of $\alpha \equiv A/(N l_P^2)$ could be inferred from measurement on BH quasinormal modes \cite{maggiore2008}.

\subsection{Topological phases and the emergence of space from evaporating BHs}\label{SubSecEmergingSpace}

How is the existence of different phases of matter compatible with the finiteness of degrees of freedom? Such issue is closely related with the evasion of Stone-von Neumann theorem~\cite{stone1930,neumann1931,umevit1985}. In fact, it is known that in quantum mechanics all continuous, irreducible representations of Weyl--Heisenberg (for bosons) or Clifford (for fermions) algebra, are unitarily equivalent. However, as it was previously reminded, such theorem does not apply to QFT, where systems with an infinite number of degrees of freedom are studied~\cite{friedrichs1953,barton1963introduction,itzykson2012quantum,smaldone2017}. The existence of unitarily inequivalent representations of canonical (anti)commutation relations permits to describe transitions among disjoint phases of the same system, in the QFT framework.

However, it is known that it is also possible to evade the Stone--von Neumann theorem by relaxing the continuity hypothesis~\cite{strocchi1993}. This has been shown in quantum mechanical systems with a multiply-connected configuration space~\cite{Schulman1968,schulman2012techniques,kastrup2005} or in polymer quantum mechanics~\cite{ashtekar2003,Corichi2007,polymer2021}.

In Ref.~\cite{vortex} the authors studied an example where both \emph{thermodynamical} and \emph{topological} disjoint phases are realized: a vortex solution in a QFT with a spontaneously broken $U(1)$ symmetry was analyzed by means of the \emph{boson transformation method}~\cite{umevortex1975,umeExtNonAb1981,blasone2002}. Such idea, firstly developed by H. Umezawa and collaborators, permits to describe classical extended objects emerging from an underlying QFT, by means of a canonical transformation performed on bosonic quasi-particle fields, which induces an inhomogeneous condensate on vacuum. Then, the authors showed that spontaneous symmetry breaking (SSB) is indeed possible even when the volume of the system stays finite \cite{vortex}. This represents a first step to understand the emergence of different phases in the $X$ons model.

The above method also permits to shed some light on the mechanism of formation of space and quasi-particles from an underlying $X$ons dynamics. In Ref.~\cite{XonicGravity} the authors face the delicate and fascinating issue of how space itself might be viewed as a classical extended object stemming from the SSB of underlying quantum dynamics, with the associated Goldstone bosons. In that case discrete $X$ons $\Psi_j$ are approximated by a field $\Psi(x)$, and the space structure and geometric tensors (metric, curvature, torsion) emerge as a result of the condensation of Goldstone bosons, while quasi-particles are described by fields on a classical (curved) space.

\section{Concluding remarks and future perspectives of the graphene analog enterprise}\label{SecConclusions}

QG and other fundamental scenarios can be tested also with analog experiments. In fact, the exciting and rapidly evolving field of analog physics is facing a new era. The interest is shifting from the reproduction of the \textit{kinematical} aspects of the Hawking/Unruh phenomenon, that has reached a climax of precision and accuracy, to the realization of some form of BH (thermo)dynamics. The latter is a challenging problem, but given its importance, even a partial solution is surely worth the effort.

The primary goals of the research in this field should then be: to search for realizations of such dynamical aspects, and to learn from the above on QG. Here we have described the results found following the road of graphene. Let us now collect the many directions we see departing from there.

\subsection{Hunting for analog BH (thermo)dynamics} A conservative approach to BH evaporation~\cite{page1,page2} assumes that the evolution of the collapsing matter to produce a BH and its subsequent evaporation is a unitary process. This is what we would like to test in our analog systems. Indeed, current ongoing work \cite{XonicGravity} primarily focuses on the emergence of space in a QG scenario, as described in Subsection \ref{SubSecEmergingSpace}, henceforth from there we are learning on the hunt for a BH dynamics on graphene and other DMs. In fact, the results of that general work will help us construct an experimentally sound geometry/gravity theory that describes the dynamics of the elastic DM membrane and explore the relations to existing gravity models. Having an action, we would be able to compute the Wald entropy, in the usual way~\cite{Wald1993}.

With this in mind, we are studying the realization of BHs on DM, based on the discoveries we have summarized in the Subsection \ref{SubSecRamifications}. One important case under scrutiny is that of the BTZ BH realized using hyperbolic pseudospheres~\cite{GUPBTZ}. We shall operate through theoretical investigations but will interact more and more with the experimentalists to test the formulae obtained in~\cite{iorio2012,Iorio2013} (or variants, obtained by refining earlier computations in the
light of the new results, see, e.g., \cite{ip3}), and we shall produce more predictions of this kind for different samples' morphologies, and various graphene observables.

In the ``time-wise'' approach, the focus will be on reproducing BH's (and other non-trivial) emergent metrics, by suitably engineering the interaction of the electromagnetic field with the appropriate DM. The basis for the study are two kinds of results obtained in earlier investigations and discussed here. On the one hand, the emergence of the Hawking/Unruh effect for specific spatial geometries. On the other hand, the great level of accuracy reached with laser pulses to control spatial and temporal resolution for graphene's electrons dispersion relation~\cite{lasergraphenePRL2018}. The latter results have inspired Ref.~\cite{ip4}, where important details are obtained that will pave the way to a full understanding of how to engineer suitable temporal components of the emergent metric, and how to control their dynamics. The two approaches are, of course, tightly related, as one goal will be to rephrase the spatial analysis of previous work into a temporal language, namely by identifying the appropriate transformations among spatial and temporal nontriviality of the emergent metric, and by envisaging the physical setups that could realize those metrics in a laboratory, for instance in the laser-DM interaction.

On the more proper QG side, we expect that lattice effects will play a role even within the continuous approximation regime~\cite{ipfirst}, but surely at the ``very high energy'' regimes, where the linear approximation no longer works, these aspects become dominant. In the latter regimes the (pseudo-)relativistic structure of the Dirac field will be deformed, and the discrete nature of the space(time) becomes so important that the continuum description, in terms of smooth metrics, will no longer be valid. This will become an important point to enforce analogy with QG scenarios of the discrete spacetime. Indeed, the results in~\cite{egypt2018}, where the natural analog of the Planck length is the lattice spacing $\ell$ of the material, point in that direction.

\subsection{BH entropy, the information paradox and the Xons model} Having in our hands a suitable emergent gravitational dynamics, along the lines of what explained in Subsection \ref{SubSecEmergingSpace}, it surely will be a great advancement and a necessary step towards analog BH thermodynamics. Still it would not be enough, as a suitable and reliable analog of a BH entropy is the key problem to be solved. In this respect, we have two roads in mind, one easier than the other: \textbf{i)} Entanglement entropy of the Dirac fields, on the given dynamical emergent BH background; \textbf{ii)} Computation of the Wald entropy through standard classical calculations based on the experimentally sound geometry/gravity theory that describes the dynamics of the elastic graphene membrane.

The first approach is easier, in two respects. First, one does not need an action for the geometry/gravity theory in point. Second, there are many results at our disposal on the entanglement entropy, from the general ones on generic bipartite systems~\cite{nielsen2000quantum}, to the specific ones on the BH thermodynamics~\cite{tera,Maldacena:2001kr,alfgae}. With these in our hands, we can surely attempt various things, and it will be exciting to see how certain issues of the theoretical side are solved here in practise. Given the results on the granular regimes beyond the linear theory~\cite{egypt2018,ip5} (see also~\cite{cagliari}), we are also in the position to compute QG corrections to the formula, and compare theoretical predictions on the QG side, as well as experiments on the condensed matter side. The second approach is more difficult, nonetheless we plan to also move steps in that direction because of its more direct link with purely gravitational scenarios. An exciting perspective is that these two approaches are complementary. It will be illuminating to compare the two, Bekenstein/Wald and entanglement.

To have these aspects under control is clearly necessary to face, within this approach, the long-standing issue of the information loss. In~\cite{aischol} it was investigated the impact on the Page curve of a picture born in analogy with condensed matter, named there the ``quasi-particle picture''. In this picture, more fundamental entities exist (we might call them $X$ons, with Feynman), and they make particles and spacetime at once: hence the (information preserving) unitarity of the BH evaporation of the Page curve, is not tenable. In~\cite{aismal} it is shown how entanglement, Bekenstein and thermodynamic entropies all stem form the same operator, whose structure is the one typical of Takahashi and Umezawa's TFD~\cite{TFD2}. We expect that the several interesting new insights gained from this work, will substantially help to reach the goals.

Finally, in~\cite{GUPBTZ}, taking advantage of the peculiarities of the BTZ BH~\cite{BTZ1992}, the extremal $M=0$ case was identified as furnishing an alternative way to the emergence in DMs of a maximal resolution/minimal length, given by the lattice length $\ell$, and related to the (negative) cosmological constant as $\ell = 1 / \sqrt{-\Lambda}$. Noticeably, a similar independent proposal emerged in the discussion of the entanglement entropy of the BTZ, see~\cite{cadoni2010}. There, the AdS length is promoted to the typical length below which spatial quantum correlation are traced out. Clearly, this road has the potential to produce very interesting results.

\subsection{Other hep-th scenarios on DMs} Many other aspects, that will contribute towards the main goal of testing QG scenarios with DMs, but that are important on their own right, are in sight. Let us mention one.

The action of graphene can be recast \cite{ip3} in a form very similar to the action of USUSY, for an external non-abelian $SU(2)$ gauge field and a fixed curved background \cite{AVZ}. Indeed, if the geometric background is fixed and the non-Abelian gauge field is external (there is no dynamics for the phonons and gauge fields), then the only difference between such actions is the coefficient in front of the torsion term. Interestingly enough, the vacuum sector is defined by configurations locally Lorentz flat, as is the case of BTZ BHs~\cite{BTZLorentzflat}, and $SU(2)$ connections carrying nontrivial global charges \cite{USUSYSU2}.

\subsection{HELIOS} Let us close by making the case for a laboratory where QG and other fundamental theories of nature are tested with analogs. Having in mind what we discussed at length in this review, and that these are the days of the AdS/CFT correspondence (see, e.g., \cite{susskindBOOK}), relating gravity and matter, we believe that the times are mature for a dedicated laboratory, entirely devoted to test fundamental theories by using analogs~\cite{helios1,helios2,helios3}. A laboratory built with the same spirit of CERN will unify, systematize and organize those efforts, but will also raise the status of the analog enterprise to a quest to reach beyond the known. The other side of the story is that analogs are often important materials for technological applications, like the case of graphene discussed in this review. Such a laboratory would then be an invaluable think-tank, where unconventional thinking would be routinely applied to create new technology, and to solve fundamental problems. It is a long while that, within our research group in Prague, we call this future facility \textit{HELIOS}, for High Energy Lab for Indirect ObservationS.

\section*{Acknowledgements}
We gladly thank Lello Agostino, Francisco Correa, Arundhati Dasgupta, Gaston Giribet, Paco Guinea, Vit Jakubsky, Siddhartha Sen, Guille Silva, Maria Vozmediano and Jorge Zanelli, for the many stimulating and informative discussions on the experimental and theoretical physics of this ``graphene analog enterprize''. L.~S. acknowledge the kind hospitality of the Institute of Particle and Nuclear Physics of Charles University.
This research was funded by Charles University Research Center (UNCE/SCI/013), by Fondo Nacional de Desarrollo Cient\'{i}fico y Tecnol\'{o}gico--Chile (Fondecyt Grant No.~3200725) and by the Polish National Science Center grant 2018/31/D/ST2/02048.

\bibliographystyle{ieeetr}
\bibliography{Bibliografia_Universe}

\end{document}